# Modular Metamaterials for Adaptable MRI Signal Control: Combining Dipole Arrays with Hexagon-based Artificial Dielectrics

Santosh Kumar Maurya [1,2], Ilan Goldberg[3], Rita Schmidt[1,2]

[1]Department of Brain Sciences, Weizmann Institute of Science, Israel

[2]The Azrieli National Institute for Human Brain Imaging and Research, Weizmann Institute of Science

[3]Department of Neurology, Rabin Medical Center, Israel

**Abstract**

Emerging concepts of metamaterials for magnetic resonance imaging (MRI) offer improved image quality through enhanced RF field control. However, current designs are often limited by bulky configurations, dielectric losses, and limited adaptability. Patient-specific optimization—essential for improving field homogeneity, boosting local SNR, and addressing anatomical variability—requires redesigning each setup, including capacitor values and structure dimensions. We present two strategies for overcoming these limitations. First, we introduce a hexagonally-packed artificial dielectric, free of any lumped-elements, capable of achieving a wide range of high relative permittivities ($\varepsilon_r \approx$ 30–1400) suitable for clinical and ultra-high-field MRI. We identify multilayered and in-plane-shifted hexagonal configurations, that optimize field interactions, supporting compact modular layouts. Second, we combine these artificial dielectrics with dipole arrays to create a modular system for adaptable signal enhancement, tuned via electric and magnetic dipole modes by adjusting the length of conductive strips. In-vivo 7T MRI results demonstrate enhanced field control and high-resolution imaging. This lightweight, flexible, and modular platform enables patient-specific RF field shaping, thus opening up new avenues for personalized and adaptable MRI technology.

## Introduction

Magnetic resonance imaging (MRI) has become a cornerstone of modern medical diagnostics, with over hundreds of million examinations conducted annually worldwide. Advances in parallel imaging using multi-channel receive coils and the development of ultra-high-field MRI systems (≥7 Tesla) have enabled sub-millimeter, three-dimensional imaging of the human brain[1], further advancing the promise of personalized medicine. Yet, MRI is an approach inherently limited by its signal-to-noise ratio (SNR), always in need of higher SNR to enable higher spatial resolution and shorten scan duration. Shorter scan durations are paramount for both patient comfort and better image quality, as they offer greater resilience to motion artifacts.

Higher SNR is typically achieved by placing the receive coils as close as possible to the patient, thereby maximizing the filling factor. However, standard coil arrays are designed with fixed geometries and patient comfort in mind, leading to suboptimal configurations for individual anatomies. Recent wireless technologies[2–7]—including dielectric pads and metamaterials—have demonstrated localized SNR improvements in brain, body, and functional MRI.

In addition to reception, metamaterials concepts can contribute to improving transmission[8–10]. MRI relies on spin-excitation induced by the radiofrequency (RF) transmit field, commonly referred to as $B_1^+$, which determines the flip angle according to $|\gamma B_1^+ \tau|$, where $\gamma$ is the gyromagnetic ratio and $\tau$ is the RF pulse duration. Inhomogeneity in the $B_1^+$ field can lead to local signal loss and degraded diagnostic contrast. Such RF field inhomogeneity arises when the wavelength at the corresponding RF Larmor frequencies[11–13] becomes comparable to the dimensions of the scanned object —a challenge encountered during abdominal imaging at 3T and the imaging of any body part at ≥7T. Although much progress has been made to provide MR hardware solutions for this issue, especially multi-channel transmit coils[14–20], flexible and local solutions can offer localized alternatives.

Dielectric materials added as pads to the MRI setup have previously been shown to improve RF homogeneity[21–24], increase local SNR[25–30], and reduce power deposition[21,26,31]. However, such setups are typically bulky and subject to significant dielectric losses. To address these limitations, metamaterials have emerged as promising alternatives[4,32–36], offering compact and wireless means to improve RF field control. Recently, metamaterials composed of *artificial* dielectrics[37–39] have been shown to provide thinner and lighter solutions. Nevertheless, these designs rely on lumped capacitive elements, which increase fabrication complexity and contribute to additional losses.

A major challenge of current metamaterial setups is that once fabricated, their electromagnetic (EM) properties are fixed. Patient-specific optimization—essential for improving field homogeneity, boosting local SNR, and addressing anatomical variability— requires redesigning each setup, including capacitor values and structure dimensions. As a consequence, there is a growing demand for metamaterials capable of tailoring local SNR enhancement in a patient-specific manner.

In this study, we present two strategies to overcome these constraints. First, we develop an *artificial* dielectric free of lumped elements, based on hexagons having a high packing efficiency (a feature frequently exploited in natural systems). Building on recent capacitive-grid-based dielectrics[37], we propose a structure comprising two shifted networks of conducting strips, on different sides of a dielectric substrate, and with each having sub-wavelength gaps arranged in a hexagonal pattern (see Fig. 1). The hexagonal pattern increases the effective capacitance density and enables high relative permittivity suitable for clinical and ultra-high field MRI applications. We further explore multilayer and in-plane-shifted configurations to enhance EM field interactions and support more compact layouts. These principles enable easy implementation of thin, flexible metamaterials, with the multi-layer configurations offering a modular approach for tailored signal enhancement.

Second, we augment these artificial dielectrics with arrays of conducting strips to construct reconfigurable metamaterials. By varying strip lengths, we control electric and magnetic dipole resonances, enabling tunable signal enhancement in specific regions of interest. Our previous work[40] analyzed the contribution of an array of electric and magnetic dipoles in a hybrid design that combined dipole arrays with a high permittivity dielectric material, demonstrating that the transmit RF magnetic field in MRI can be tailored. Here, we combine the new flexible *artificial* dielectric with an array of electric and magnetic dipoles and demonstrate its benefits for in-vivo imaging. This design offers modularity by enabling easy switching between configurations, thus easily controlling electromagnetic properties and signal enhancement, paving the way for patient-adaptable applications. We demonstrate its benefit in MRI applications relevant for clinical and research studies.

## Results

### Hexagonally packed artificial dielectric – single, multi-layer and in-plane-shifted configurations

To achieve compact subunit designs for artificial dielectrics (AD) suitable for MRI applications, we identified a hexagonal packing configuration capable of providing the desired range of relative permittivity. In this design (Fig. 1B), six conductive three-blade-propeller shaped subunits are arranged to form a hexagon. Three subunits are laid on one side of the dielectric substrate, interleaved with three on the opposite side, creating a layered structure. A beehive-like assembly of such hexagons forms a distributed capacitive network. For benchmarking, a conventional rectangular layout (Fig.1A) —similar to previously reported designs[37–39]—was assembled. The hexagonal arrangement enables the formation of an effective capacitor network without the need for discrete lumped components, thereby simplifying fabrication and enhancing the achievable effective permittivity.

**Single-layer Hexa.** In this study, the hexagonal artificial dielectric (AD) was designed for brain imaging applications; therefore, the overall dimensions were selected to accommodate integration within a standard RF head coil setup. The configuration shown in Fig. 1 was designed with a footprint of 24 × 14 cm² accordingly. As a first step, a hexagonal layout was designed (Fig.1B) and compared to rectangular grid-of-crosses layout (Fig.1A). The hexagonal configuration achieved a relative permittivity of $\varepsilon_r \approx 80$, a twofold increase compared to the conventional grid-of-crosses layout ($\varepsilon_r \approx 40$) of the same overall dimensions. The effective permittivity values reported for the artificial dielectric above were based on simulations using 400-micron-thick dielectric substrate with $\varepsilon_r$=2.6, selected to demonstrate feasibility with readily available materials—corresponding to the thickness of commonly available plastic sheets. This approach supports low-cost prototyping and broader accessibility for practical implementations. Additionally, with flexible PCB printing technologies, substrates of thicknesses of 25,50,100,200 microns can be utilized, enabling the fabrication of ultra-thin and conformal artificial dielectrics. As relative permittivity increases with thinner dielectric substrate, the permittivity of this flexible PCB setups increases 16-fold when using a 25 microns substrate (compared to the 400 microns above). This range of substrate thickness yields an overall relative permittivity values of 80 to 1280 for the hexagonal configuration. Different PCB materials can be also utilized to further increase effective permittivity.

**Multi-layered Hexa.** Another configuration of interest is an artificial dielectric designed for smaller overall dimensions—for example, to achieve localized enhancement in the auditory region of the brain. To maintain the same relative permittivity as the larger design ($\varepsilon_r \approx 80$) with a comparable number of subunits, we explored the potential of multilayered structures. A triple-layered hexagonal design (Fig. 1C) was configured with overall dimensions of 17 × 11 cm² (with ~two-times smaller area than in Fig.1B), where each layer was laterally shifted to create overlapping regions, thereby increasing the overall capacitive network. This stacking approach yielded a 1.8-fold increase in permittivity for a double-layered setup, and a 2.5-fold increase for the triple-layered configuration, while maintaining the same footprint (Fig. 1C). This compact multilayered design offers three key advantages: (1) higher permittivity within the same physical dimensions, (2) support for smaller layouts without compromising the operational frequency range, and (3) a modular architecture that allows the tuning of the overall permittivity based on patient-specific anatomy and/or on application.

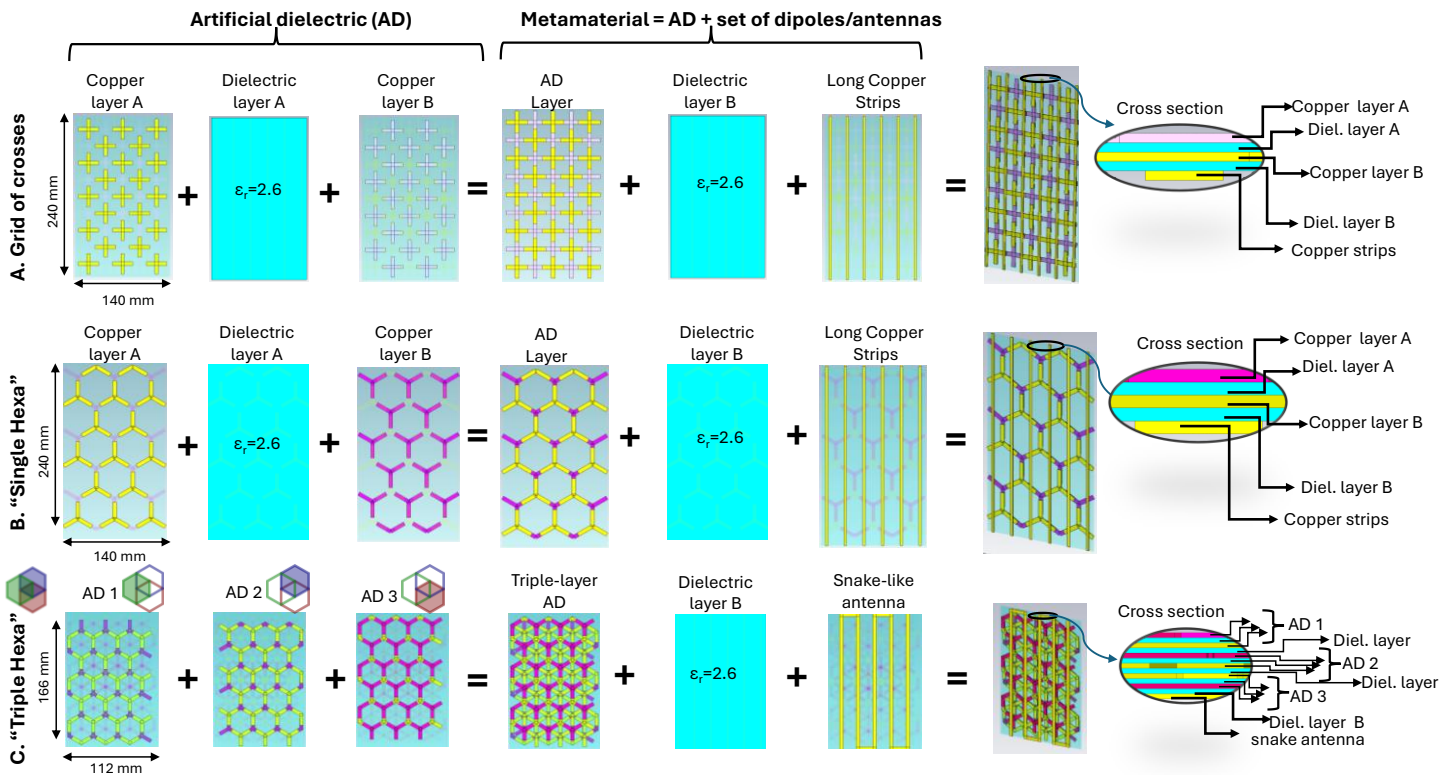

**Figure 1. Schematic metamaterial design.** Top to bottom shows A) "Grid of crosses", B) "Single-Hexa", C) "Triple-Hexa" configurations. From left to right the schematics of the AD is shown and then the combined setup of the metamaterial. A cross section of the layers is included. The "Triple-Hexa" AD includes three AD layers, where two layers are diagonally shifted and attached to the first one. The shift between the layers is depicted by the picture of three hexagons (green, blue and red).

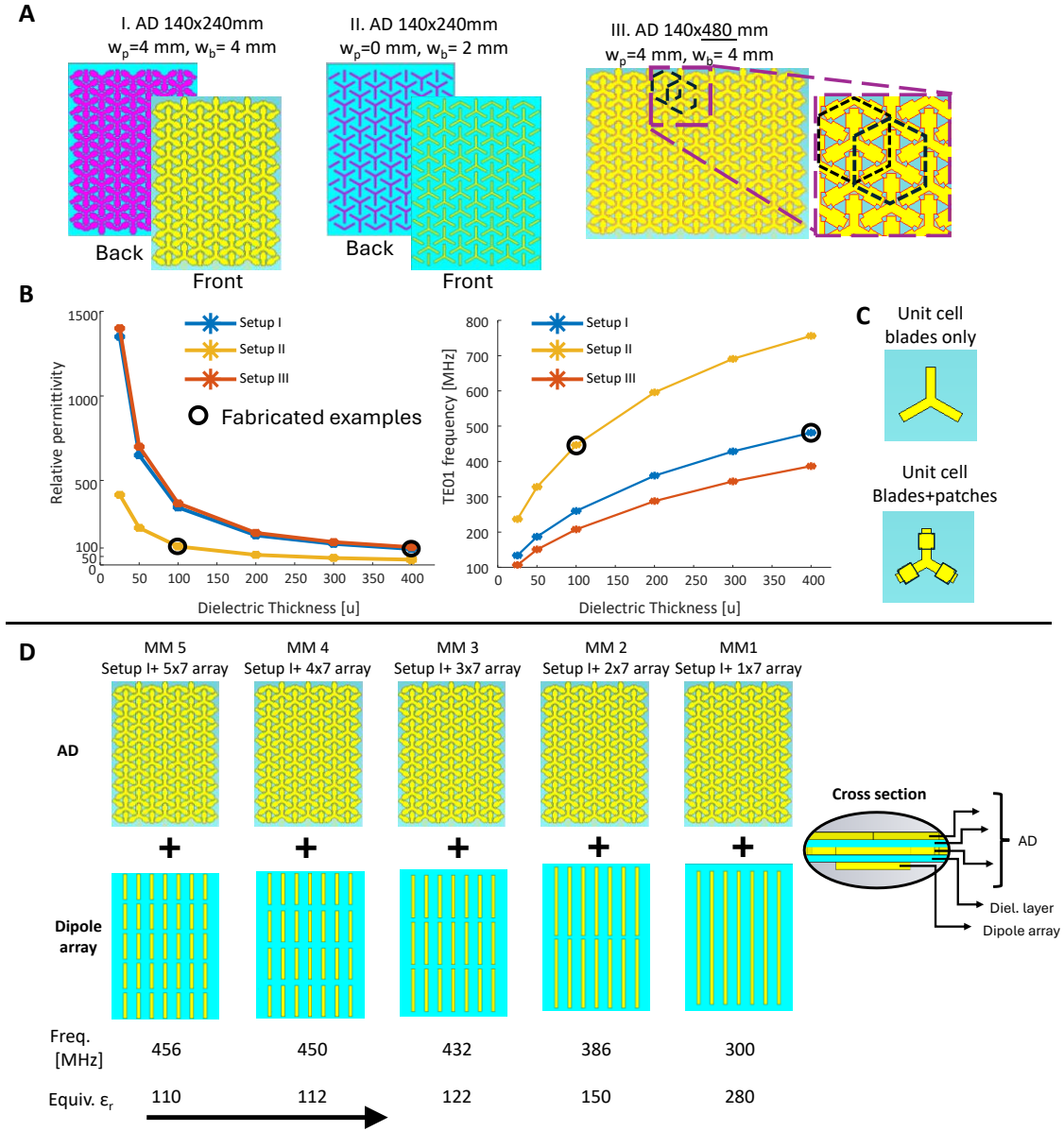

**Figure 2. Combining artificial dielectric and dipole arrays. A)** In-plane shifted hexagonal configurations. Simulations for three setups are shown: I) Layout with subunits containing blades and patches (using a 400-micron dielectric substrate with $\varepsilon_r$=2.6, yielding $\varepsilon_r$ = 95);II) Layout with subunits containing blades only (using a 100-micron dielectric substrate with $\varepsilon_r$=2.6, yielding $\varepsilon_r$ = 95); III) Setup with twice the in-plane dimensions compared to I and II. **B)** Relative permittivity ($\varepsilon_r$) and $TE_{01}$ mode frequency as a function of dielectric substrate thickness for the three layouts shown in A. **C)** Schematics of a single unit cell with and without patches. **D)** Modular setup combining the artificial dielectric with dipole arrays of varying lengths. Below each configuration, the corresponding $TE_{01}$ frequency and equivalent $\varepsilon_r$ (calculated for a homogeneous dielectric with the same frequency, in-plane dimensions, and 7 mm thickness) are indicated.

**Single layer in-plane-shifted Hexa.** Since multilayered designs increase the overall thickness, an alternative approach using in-plane shifted hexagons was examined to further increase the effective permittivity (Fig. 2). Here we successfully achieved $\varepsilon_r \approx 100$ with 400-microns thick dielectric substrate (Fig.2A, setup I), which is a value previously shown[29,41] to be effective for brain applications using conventional dielectrics at 7T MRI.

To provide control over the resonance frequency and effective relative permittivity, the design incorporated the following parameters: blade width ($w_b$), patch width ($w_p$), and dielectric substrate thickness. Figure 2A (Setups I and II) illustrates configurations with and without the added patches, respectively. Additionally, the setup was tested with overall dimensions twice as large (Fig. 2A, setup III), making it more suitable for body imaging applications. Considering the dielectric substrate thicknesses supported by flexible PCBs, the examined setups achieved relative permittivity values ranging from 30 to 1400 (Fig.2B). This broad range supports diverse applications—including brain[22,28,29,42] and body imaging[31,43,44]—across MRI field strengths with relevant applications for 1.5T MRI and above, with a particular interest at 3T and 7T.

To compare the performance of artificial dielectrics to conventional dielectrics, we analyzed the lowest transverse electric ($TE_{01}$) mode for each configuration. The effective permittivity of the artificial dielectric was defined as the relative permittivity of a 0.7 cm thick conventional dielectric, sharing the same in-plane dimensions and $TE_{01}$ resonant frequency (Fig.2, Fig.3). The comparison was made to a 0.7 cm thick conventional dielectric, since it is a commonly used size in MRI applications. The $TE_{01}$ mode is particularly relevant for MRI applications, as it generates magnetic field (H-field) components perpendicular to the main static magnetic field ($B_0$) and exhibits a large penetration depth. Critically, this mode also minimizes the electric field within the subject or in regions external to the structure, thereby reducing the potential for unwanted power deposition. The distribution of the $TE_{01}$ mode of the artificial and the conventional dielectrics exhibited similar distributions.

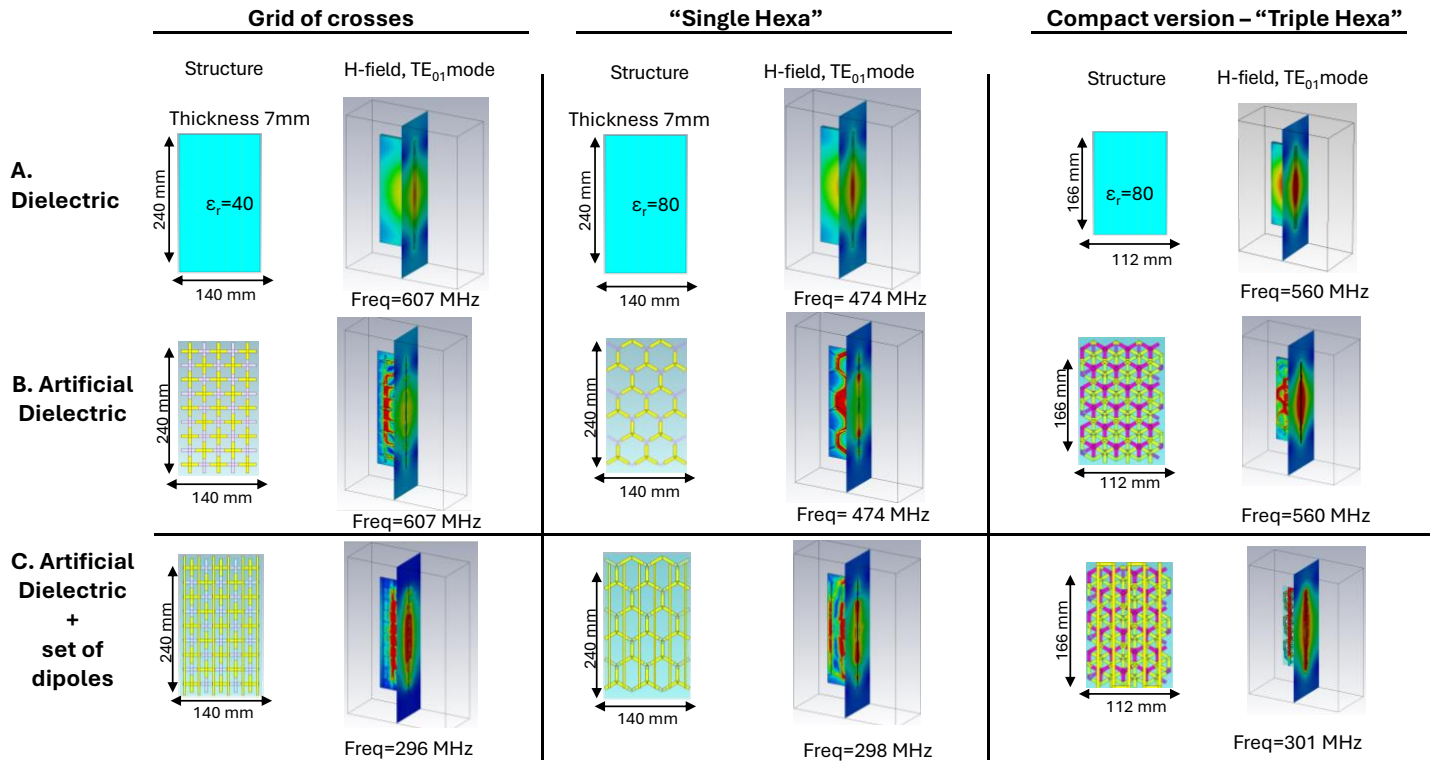

**Figure 3. H-field maps of AD and metamaterial configurations**. The H-field of the $TE_{01}$ mode for each AD and the metamaterial configuration is shown. The AD H-field is also compared to a 0.7 cm thick dielectric layer with the same resonant frequency. All three metamaterial configurations were designed to generate a $TE_{01}$ resonance at ~298 MHz. Each map was scaled separately to optimally visualize the distribution.

## Combining Dipole Arrays with Artificial Dielectrics for Adaptive Signal Control

While artificial dielectrics alone can offer a wide range of tunable EM properties, each configuration typically requires a dedicated printed layout. To address this, we investigated a hybrid approach that combines artificial dielectric structures with an array of conducting strips, where the added strip array introduces additional control over the EM field distribution. Similar combinations with conventional dielectric materials have been previously investigated[4,32,35,40], but here we extend this concept to artificial dielectrics and increase design flexibility.

For each configuration shown in Fig. 1, an array of long conducting strips was added on top of the artificial dielectric, separated by a thin dielectric substrate. As an initial test, strip lengths were tuned to set the $TE_{01}$ mode resonance frequency to the 7T Larmor frequency (298 MHz). The base artificial dielectric structure—without the strip array—was intentionally designed to be non-resonant at 7T, providing a local $B_1^+$ enhancement in the range of 10–40%.

While an array of long conducting strips (≥16 cm) generates electric dipole resonances, an array of short conducting strips (e.g., 3 cm, where 3cm ≪ λ, and λ≈12 cm in brain tissue at 7T MRI) behaves analogously to an array of magnetic dipoles—functionally similar to cut split-ring resonators (Ref. [45]). This configuration also supports a $TE_{01}$ resonant mode, although it appears at a higher frequency compared to the long-strip setup. To explore the RF field control with this hybrid system, we combined the AD structure shown in Fig. 2A setup I with arrays of conducting strips of varying lengths, ranging from 3 cm to 16 cm (Fig.2D). The longest strips (16 cm) correspond to configurations dominated by electric dipole behavior, while the shortest strips (3 cm) exhibit characteristics of magnetic dipoles. Intermediate lengths provide a gradual transition between these two regimes, enabling continuous control over the EM response of the system. Fig.2D demonstrates these effects, showing that substantial control over the effective EM properties—and, consequently, the local RF field enhancement—can be achieved by modifying only the conducting strip array. All this while retaining the same underlying artificial dielectric structure, a structure which is more complex and time-consuming to fabricate than simple strips.

By preparing the two components separately—the artificial dielectric and a set of interchangeable conducting strip arrays—the system can easily be reconfigured and adapted to suit specific applications, anatomical targets, or field strength requirements without the need for reprinting the entire structure.

To evaluate the performance of the above metamaterial configurations with varying strip lengths, experimental measurements with a rectangular phantom at a 7T MRI were performed. The results demonstrate that the maximum local $B_1^+$ enhancement varied depending on the chosen configuration, reaching 10%, 14%, 22%, and 70% for setups AD1, MM3, MM2, and MM1, respectively (Fig. S1). A subset of these configurations (AD1 and MM2) was further tested using a 3D head-shaped phantom[46] at 7T to better mimic realistic imaging conditions, yielding results consistent with those observed in the rectangular phantom experiments (Fig.S2). Additionally, scans at a 3T MRI were performed using the MM1 configuration (Fig. S3), as it features $\varepsilon_r \approx 300$—a value previously shown[43] to be effective in brain imaging at 3T. In this case, a 50% increase of the maximum local $B_1^+$ was observed, confirming the adaptability and effectiveness of the proposed design across different magnetic field strengths.

## Adaptive metamaterials configurations for in-vivo applications

Several brain regions—including the occipital lobe, cerebellum, and brainstem—are particularly susceptible to local signal dropouts in 7T MRI due to RF field inhomogeneities. Head size plays a significant role in these effects: larger head dimensions intensify destructive interference patterns, resulting in more pronounced signal loss, particularly in the posterior and temporal lobes. To investigate this phenomenon, EM simulations were conducted using the virtual human model "Ella"[47], with head dimensions scaled to 0.9×, 1.0×, and 1.1× of the original size. Simulations show that in the larger head model (1.1×), the maximum $B_1^+$ at the center of the head is 14% lower compared to the smaller head (0.9×), while the minimum $B_1^+$ in the displayed slice is reduced by a factor of five.

To enhance $B_1^+$ intensity in the posterior brain region (containing the visual cortex) metamaterial configurations (artificial dielectric alone and MM1–MM4) were evaluated for their ability to provide local enhancement when placed as a thin pad at the back of the head. While phantom results showed that different metamaterial setups produced varying levels of $B_1^+$ enhancement, in this EM simulation set, we examined which configurations would yield similar $B_1^+$ distributions across different head sizes. Specifically, the AD1 only, MM3, and MM2 configurations with, respectively, the small (0.9×), medium (1.0×), and large (1.1×) head models, resulted in a similar normalized $B_1^+$ profile (Fig.4D). Table 1 shows the maximal specific absorption rate ($SAR_{max}$) values and $B_1^+$ values at points of interest.

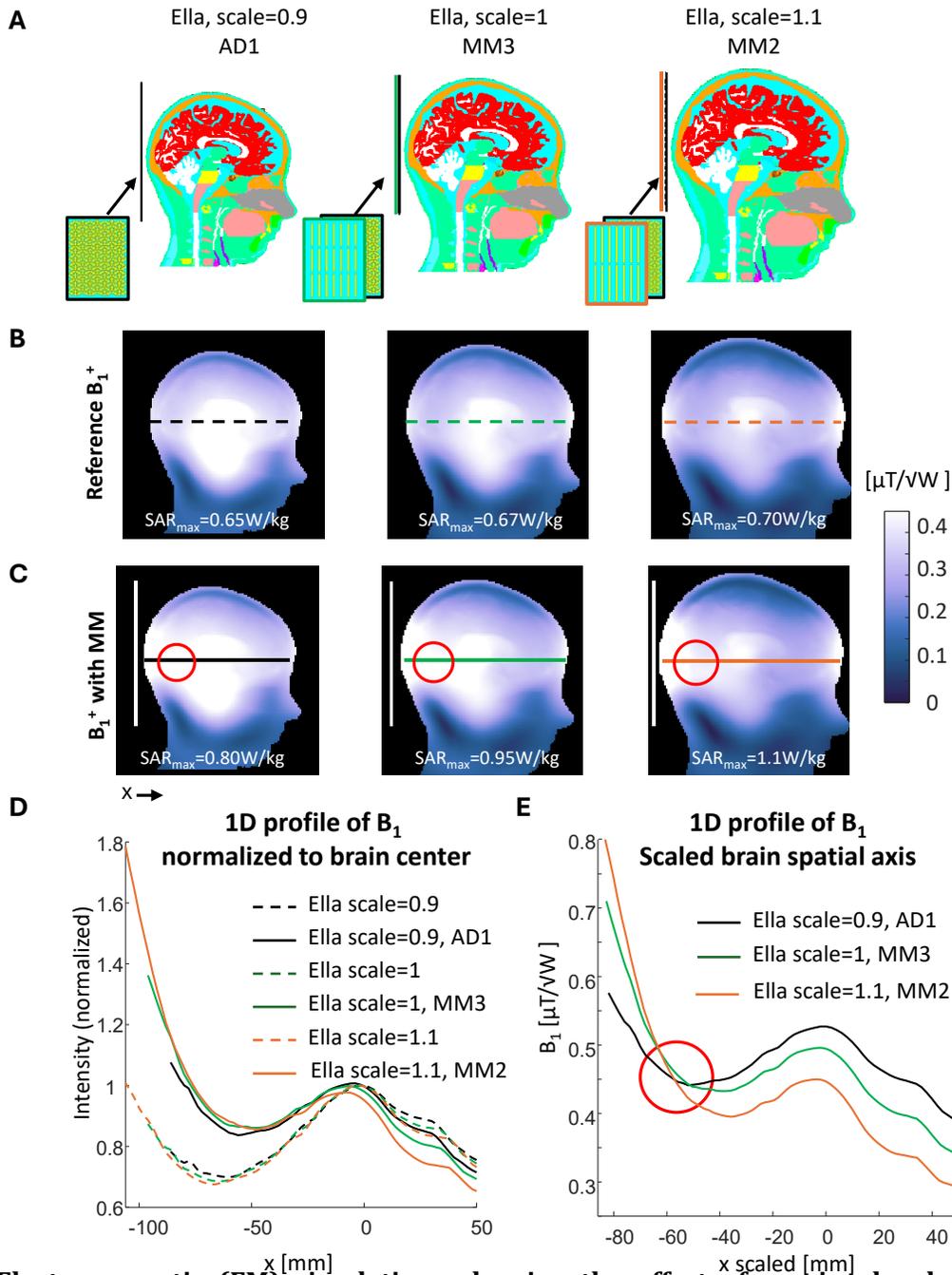

**Figure 4. Electromagnetic (EM) simulations showing the effect of varying head size and of the metamaterial configurations designed to provide similar $B_1^+$ enhancement. A)** Simulation setup, **B)** Reference $B_1^+$ maps in the central sagittal plane without the metamaterial, **C)** $B_1^+$ maps with the metamaterial in place (the metamaterial location is shown as a white overlay). **D)** 1D $B_1^+$ profile along the line indicated on the maps for each head size. **E)** 1D $B_1^+$ profile along the same line as in D, with x-axis scaled ($x_{scaled}$) to the smallest head. The red overlay circle around $x_{scaled}$=-56 mm highlights the region corresponding to the visual cortex, where similar $B_1^+$ values were achieved across all three head sizes. Specific absorption rate (SAR) values for each setup are displayed on the corresponding $B_1^+$ maps.

An important objective in MRI is achieving the same spin flip angles within the imaging region of interest across different patients. In this example, scaling the resulting $B_1^+$ profiles of the three configurations to the same size (Fig. 4E), we achieved similar $B_1^+$ amplitudes—and thus consistent flip angles—at the location of the visual cortex. For simplicity, we assumed that the position of a given brain area scales proportionally with head size. With the modular design introduced here, alternative metamaterial configurations can be selected and optimized for different anatomical targets or specific MRI applications.

Table 1: Maximum $B_1^+$ enhancement for metamaterial setups shown in Fig. 4 across varying head sizes.

|  | Ella, scale=0.9, AD1 | Ella, scale=1.0, MM3 | Ella, scale=1.1, MM2 |
| --- | --- | --- | --- |
| Ratio of $B_{1max}$, with and without MM. Each profile was normalized to the center of the brain | 1.42 | 1.65 | 1.75 |
| Ratio of $B_1$ with and without MM at x=-86 cm from the head center (corresponding to brain edge in Ella model scaled to 0.9). Each profile was normalized to the center of the brain. | 1.42 | 1.53 | 1.53 |
| $B_1$ with MM at $x_{scaled}$=-56 mm (visual cortex area) [$\mu T/\sqrt{W}$]. $x_{scaled}$ is a scaled axis matching the smallest head. | 0.45 | 0.46 | 0.45 |
| Ratio of $B_{1max}/\sqrt{SAR_{max}}$ with and without MM. Each $B_{1max}$ was normalized to the center of the brain. | 1.27 | 1.38 | 1.38 |

## Human imaging at 7T MRI

To demonstrate the potential for local signal enhancement in vivo, $B_1^+$ field maps were acquired at 7T MRI using two metamaterial configurations (AD1 and MM2), resulting in localized $B_1^+$ increases of approximately 20% and 50%, respectively (Fig.5A). To further assess the impact of the enhanced transmit field across clinically relevant applications, three MRI sequences were acquired with and without the MM2 configuration. These included: (1) a non-contrast-enhanced time-of-flight (TOF) sequence for magnetic resonance angiography (MRA), used to visualize cerebral vasculature; (2) a high-resolution 3D $T_2$-weighted sequence for detecting structural brain pathology; and (3) a multi-echo gradient-echo acquisition for generating quantitative $T_2^*$ relaxation time maps, which serve as sensitive markers of microstructural integrity and long-term pathological changes.

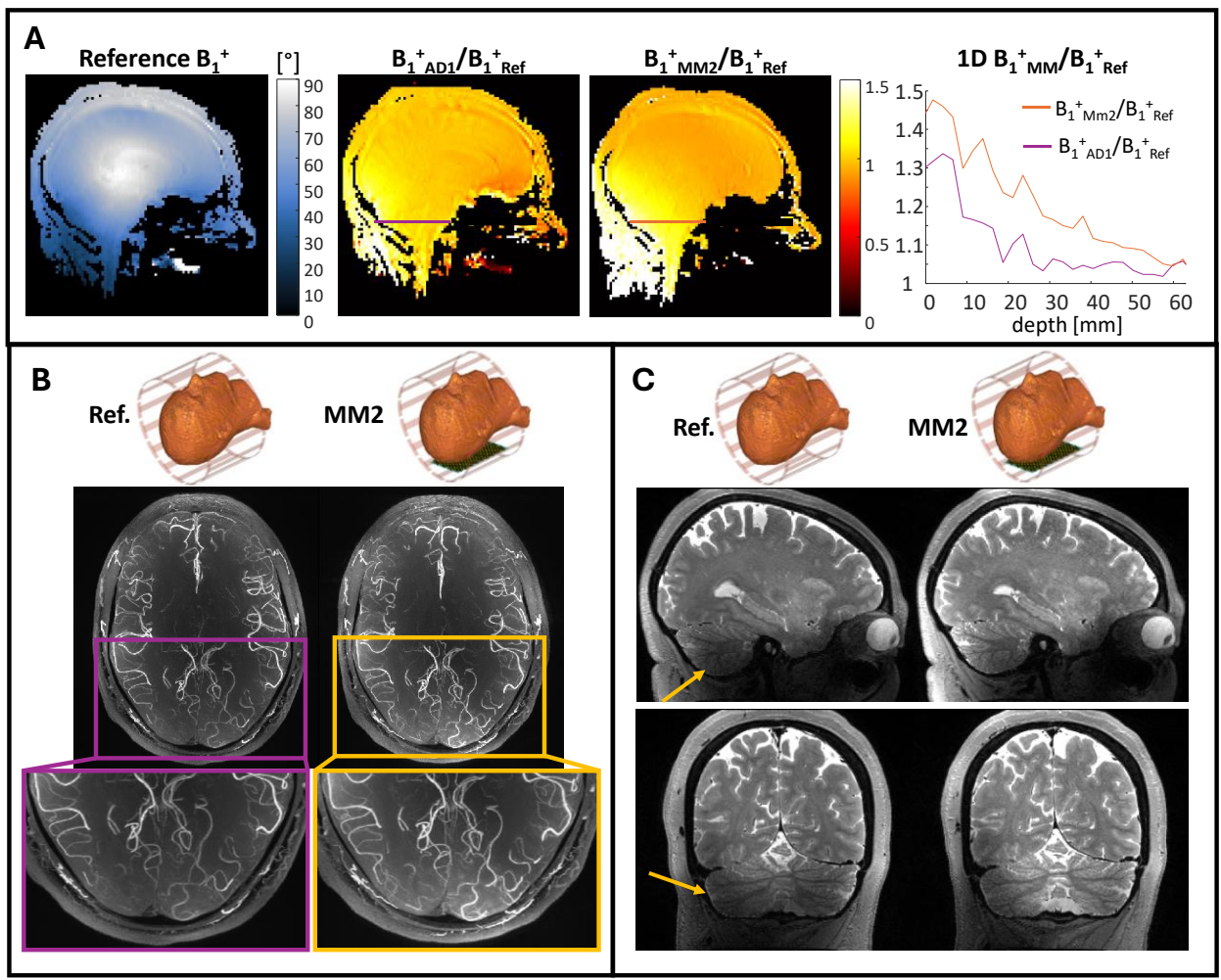

**Figure 5. In-vivo imaging demonstrating $B_1^+$ enhancement and improved image quality in posterior brain regions using an added metamaterial** A) Reference $B_1^+$ map and $B_1^+$ ratio maps with two metamaterial setups (AD1 and MM2), along with a 1D profile of the $B_1^+$ ratio along the line indicated on the maps. B) Time-of-flight angiography maximum intensity projection (MIP) images without and with the MM2 setup. **C)** Sagittal $T_2$-weighted images without and with the MM2 setup. The orange arrow highlights the cerebellum, where signal is markedly low without the metamaterial and significantly improved with the MM2 configuration.

Angiography images acquired with the metamaterial configuration showed a pronounced improvement in vessel visualization in the posterior brain region, with a two-fold signal increase in large vessels (Fig. 5B). Some of the smaller vessels were below the SNR threshold without the added structure but became clearly visible when scanned with the metamaterial in place. $T_2$-weighted scans, which are particularly sensitive to RF field inhomogeneity due to the use of refocusing pulses (with signal intensity scaling approximately as $B_1^2$), also demonstrated substantial improvement. Specifically, signal intensity in the cerebellum increased by approximately 2.5-fold (Fig. 5C).

Similarly, the multi-echo gradient echo acquisition—used to sample signal decay over a range of echo times—showed a clear signal enhancement of 1.5- to 2-fold in the cerebellum, particularly at longer echo times where signal levels are typically low (Fig. 6A). Since quantitative $T_2^*$ mapping relies on fitting the signal decay curve, low SNR in specific regions often leads to overestimation of $T_2^*$ values. By improving local signal strength, the added metamaterial enhances the reliability of the quantitative mapping (Fig. 6B–C). A cumulative distribution plot of relative fitting errors in the cerebellum revealed a 1.4-fold increase in the number of voxels with error below 10%, confirming improved accuracy in $T_2^*$ estimation.

## Discussion

In this study, we developed and demonstrated a new class of modular MRI-beneficial metamaterials combining hexagonally packed artificial dielectrics with dipole arrays. This modular platform enables local signal enhancement in MRI by tailoring the EM field in a flexible and adaptable manner.

One key advantage of the proposed artificial dielectric is the absence of any lumped elements, leveraging hexagonal capacitive networks to achieve a wide range of effective relative permitivities ($\varepsilon_r \approx 30$–$1400$), relevant for local signal enhancement at clinical and ultra-high field MRI (1.5T-7T as well as higher fields). Through multilayered and in-plane-shifted configurations, we achieved fine control over permittivity, while maintaining compact and thin layouts, making the design practical for placement between the patient and a close-fitting receive coil array. Compared to previously proposed rectangular geometries, the hexagonal configuration provided a substantial increase in effective permittivity, achieving up to a two-fold enhancement in single-layer implementations. Additional gains of up to 2.5-fold were observed with multilayer stacking. The in-plane shifted configuration was implemented using readily available materials—a 400-micron plastic sheet and 4 mm copper tape—offering a low-cost and easy fabrication approach. Additionally, a flexible printed version was developed using a 100-micron dielectric substrate, demonstrating the applicability of the design for higher-precision manufacturing processes.

Beyond control of the permittivity, the supplementing of dipole arrays adds a practical option of modularity. By varying the length of the strips, we demonstrated control over the resonant mode ($TE_{01}$), shifting the frequency from 300 to 487 MHz, which in turn modulates the level of local $B_1^+$ enhancement. This tuning, achieved by simply interchanging layouts of the strip array while keeping the artificial dielectric base fixed, allows for field-specific and anatomy-specific adjustments without redesigning the entire structure. This modularity paves the way for reconfigurable designs tailored to patient-specific clinical applications.

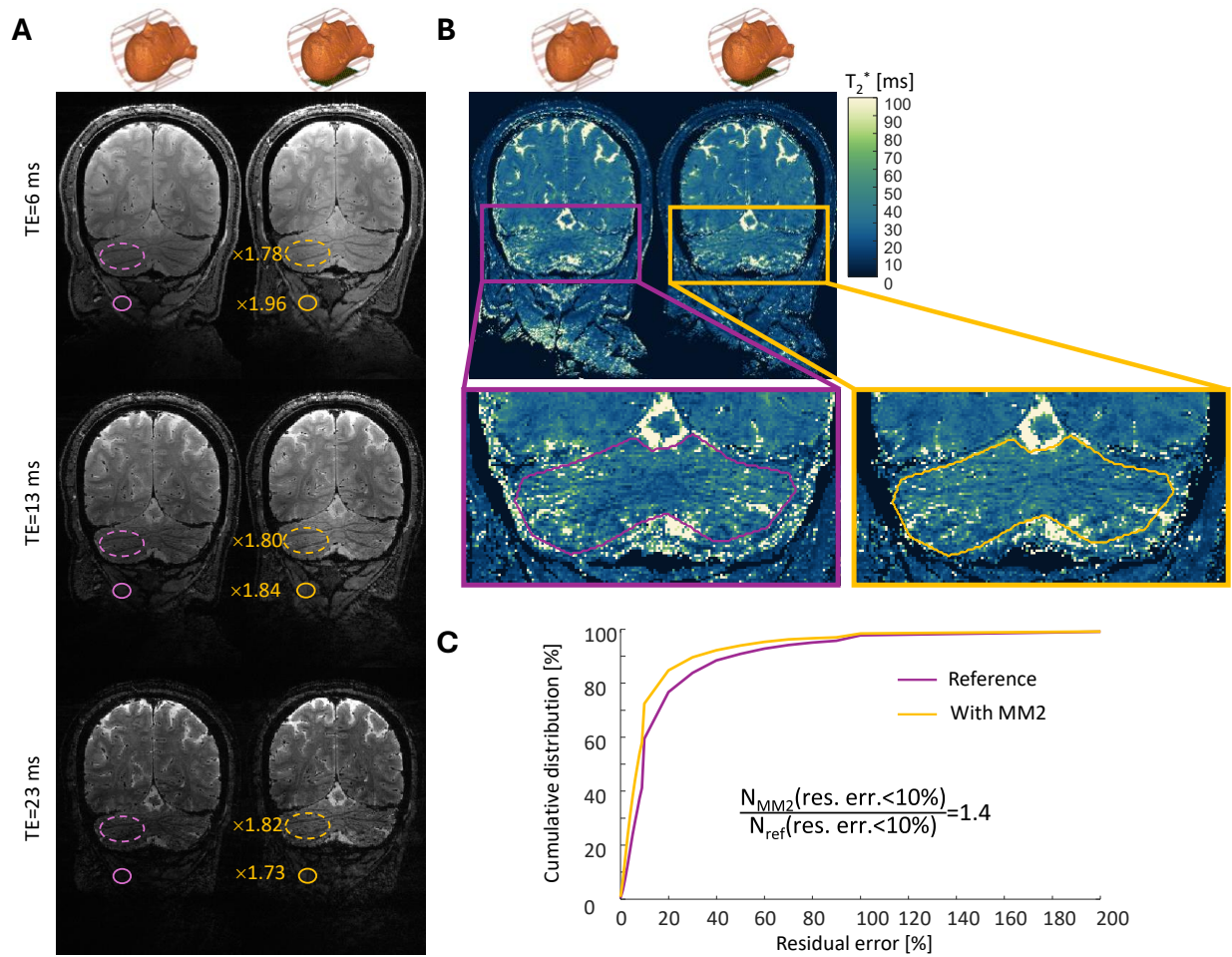

**Figure 6. Improved quantitative $T_2^*$ mapping performance in posterior brain regions using an added metamaterial.** A) Coronal magnitude images across the cerebellum acquired without and with the MM2 setup, shown for three echo times. Signal enhancement is highlighted in two regions of interest. B) Corresponding $T_2^*$ maps for the same slice, illustrating reduced overestimation artifacts in the cerebellar folds when using the metamaterial. C) Cumulative distribution of residual errors in $T_2^*$ estimation, comparing results without and with MM2.

Our simulations and phantom measurements confirmed substantial local $B_1^+$ enhancement—up to 70%—depending on the configuration. We further validated the performance using a 3D head-shaped phantom designed to mimic brain tissues at 7T and demonstrated cross-field applicability with a configuration for 3T.

To address anatomical variability, our simulations highlight the importance of selecting configurations tailored to individual head size. We showed that $B_1^+$ amplitude decreases with increasing head size, and that achieving consistent excitation flip angles across different anatomies requires adaptable tuning of the metamaterial configuration. As a representative application, we demonstrated the mitigation of posterior signal dropouts—a known challenge in ultra-high-field brain imaging —by optimizing the configuration as a function of head size.

In this work, we demonstrated applications where local signal enhancement is particularly beneficial. While MRI signal intensity depends on multiple factors—including tissue properties and pulse sequence parameters—a useful approximation can be made in the low flip angle regime to estimate SNR behavior. Under these conditions, the image SNR is approximately proportional to $\sin(\gamma B_1^+ \tau) \cdot (B_1^-) / \sqrt{P}$, where $\gamma$ is the gyromagnetic ratio, $\tau$ is the RF pulse duration, $P$ is the accepted power of the coil, and $B_1^-$ is the complex conjugate of the transmit field, defined as $(B_{1x} - iB_{1y})/2$. For small flip angles, this expression simplifies to SNR $\propto \gamma (B_1^+)^2 \tau / \sqrt{P}$, indicating that enhancements in both transmit ($B_1^+$) and receive ($B_1^-$) fields can contribute multiplicatively to the SNR. In regions affected by local $B_1^+$ dropout, improvements in $B_1^+$ may therefore yield an approximately quadratic increase in SNR—assuming a corresponding gain in $B_1^-$. However, this assumption should be applied with caution, as $B_1^+$ and $B_1^-$ serve distinct roles in excitation and signal reception, respectively, and can exhibit spatial differences. Their relative contributions depend on the specific MRI pulse sequence and acquisition strategy used. In our experiments, the observed increases in $B_1^+$ and image intensity—particularly in gradient-echo acquisitions—suggest that this relationship held for the tested configurations. These findings indicate that both RF transmission ($B_1^+$) and reception ($B_1^-$) can be locally enhanced, resulting in improved SNR without the need for active circuitry or bulky hardware. While passive setups for local enhanced SNR were previously demonstrated with dielectric pads, the new design highlights new practical aspects – light, thin and modular - for RF field shaping.

Human scans at 7T MRI further support the practical utility of our approach. $B_1^+$ mapping showed 20–50% signal enhancement in targeted brain regions, and clinical sequences demonstrated the impact of this enhancement. Angiography revealed up to two-fold signal increase in large posterior vessels, while $T_2$-weighted imaging showed clear improvement in the cerebellum—critical for assessing pathologies such as multiple system atrophy and posterior circulation stroke. The multi-echo gradient echo scans, used for quantitative $T_2^*$ mapping, benefited significantly from increased signal at longer echo times. This not only improved image quality but also enhanced the accuracy of $T_2^*$ relaxation estimates. In the cerebellum, the reliability of $T_2^*$ mapping improved significantly, with a 40% increase in the number of voxels exhibiting fitting errors below 10%.

Taken together, these findings underline the versatility, adaptability, and clinical relevance of the proposed metamaterial platform. The ability to easily reconfigure the system for different anatomical targets, field strengths, or clinical sequences, positions it as a compelling tool for personalized MRI. This solution beneficently offers low-cost compact designs that easily integrate into existing workflows.

Future work may focus on expanding the platform to automatic reconfigurable implementation as well as imaging of other body regions—such as cardiac or abdominal imaging—where RF inhomogeneity and SAR constraints are pronounced. Additionally, integration with data-driven approaches (e.g., machine learning-based $B_1^+$ mapping or real-time feedback systems) could enable fully optimized, patient-specific configurations at the point of care.

# Materials and Methods

## Characterization of the metamaterial structure

*Common design principle of the artificial dielectric:* The basic characteristic of the AD is two laterally shifted copper strip grids separated by a dielectric substrate. In our simulations and experiments we used a dielectric substrate with a relative permittivity of $\varepsilon_r$ = 2.6. This configuration forms a distributed capacitive network that determines the effective dielectric properties of the AD structure. The investigated AD designs featured hexagonal structures in single-layer and multi-layer (Fig.1, Fig.3), and in-plane shifted (Fig.2) configurations.

*Single and multi-layer AD design and EM simulations:* In Fig.1 a hexagonally structured design ('Single-Hexa') was compared to a conventional grid layout ('Grid-of-Crosses') to demonstrate more efficient spatial filling, resulting in a higher effective dielectric constant. A triple layer hexagonally structured design ('Triple-Hexa') was designed to achieve the same permittivity as 'Single-Hexa' but for smaller overall dimensions.

Electromagnetic (EM) simulations were performed using CST Studio Suite 2019 (Dassault Systèmes Deutschland GmbH), including both eigenmode analysis (Fig.3) and full-setup simulations. The eigenmode solver was used to characterize the AD structure, specifically its fundamental transverse-electric ($TE_{01}$) mode and corresponding resonant frequency. For each configuration, the effective $\varepsilon_r$ was obtained by identifying the $\varepsilon_r$ value of a conventional 0.7 cm thick dielectric yielding the same $TE_{01}$ mode frequency. Additionally, the influence of the middle dielectric layer's thickness on the AD's effective permittivity was systematically evaluated (Fig.2B).

Single-Hexa and Grid-of-Crosses configurations were each combined with a dipole array to form resonant structures at 298 MHz. Similarly, the Triple-Hexa design was paired with a snake-like antenna[48] to extend the effective electric dipole length, compensating for its shorter overall dimensions.

*In-plane shifted AD design and EM simulations:* Figure 2A illustrates AD structures based on in-plane shifted hexagonal designs, demonstrated in two sizes: 24 × 14 cm² and 48 × 14 cm². The first two configurations incorporate subunit patches to tune the effective dielectric properties. The first was optimized for low-cost fabrication, using a 400-micron dielectric substrate and 4 mm copper strips. The second was designed for flexible PCB printing, employing 2 mm copper blades and a 100-micron dielectric substrate. To achieve comparable effective permittivity with the thinner substrate, patches were included. For each configuration, the dependence of the effective relative permittivity and $TE_{01}$ mode frequency on dielectric substrate thickness was calculated.

*Combining in-plane shifted AD with dipole array:* A modular design was developed by combining the AD setup #1 from Fig. 2A with dipole arrays of varying lengths. Dipole lengths of 3, 3.6, 5, 8, and 15.6 cm were tested within a final layout of 17.4 × 12.8 cm², implemented as arrays of 5×7, 4×7, 3×7, 2×7, and 1×7 dipole units, respectively (see Fig. 2D). These configurations were referred to as MM5, MM4, MM3, MM2, and MM1, respectively. For each configuration, the corresponding $TE_{01}$ resonance frequency and the enhancement of the RF transmit field were evaluated.

*Full-setup EM simulations details:* Simulated $B_1^+$ maps were normalized to an accepted power of 1 Watt. Phantom simulations for estimating the $B_1^+$ field included a surface loop coil and a rectangular phantom mimicking brain tissue properties ($\varepsilon_r$ = 53, σ = 0.3 S/m), with dimensions of 14 × 8 × 16 cm³. Human model simulations (Fig. 4) incorporated a four-loop coil array designed to provide whole-brain coverage (Fig.S4). The overall coil coverage spanned over 34 cm in diameter and 14 cm in length.

The "Ella" human model from the Virtual Family was used with a mesh resolution of 1 × 1 × 1 mm³ to simulate the RF transmit field in the head region. To evaluate the impact of head size variability, the model's dimensions were scaled by factors of 0.9, 1.0, and 1.1. The metamaterial structure was positioned near the occipital lobe, as shown in Fig. S4.

## Metamaterials implementation

Fabricated configurations included the Single-Hexa, Triple-Hexa (Fig. 1), and in-plane shifted Hexa designs (Fig. 2I and Fig. 2II). Photographs of all setups are presented in Fig. S5. The setup shown in Fig. 1 used a dielectric substrate with relative permittivity $\varepsilon_r$ = 2.6, a thickness of 400 microns.

A flexible PCB version of Fig. 2 Setup II was fabricated by Newline-PCB using DuPont™ Pyralux® TK—a double-sided copper-clad laminate. The design featured a 100-micron-thick dielectric layer and 36-micron copper layers printed on both sides according to our specifications. Photographs of the fabricated PCB are provided in Supporting Information Fig. S4.

## MRI measurements

MRI scanning was performed on a 7T system (MAGNETOM Terra, Siemens Healthcare, Erlangen) using a 1Tx/32Rx Nova coil, and on a 3T system (Prisma, Siemens Healthcare, Erlangen). $B_1^+$ field maps were acquired using the vendor-provided $B_1$ mapping sequence, based on a preconditioning RF pulse with Turbo FLASH readout [Ref. 29], with a 200 × 200 mm² field of view (FOV) and spatial resolution of 2.5 × 2.5 × 3.5 mm³.

Phantom studies included: (i) a rectangular phantom composed of sucrose, agarose, and water, providing $\varepsilon_r$ ≈ 56, σ = 0.3 S/m (mimicking brain tissue properties) (ii) a head-shaped brain-mimicking phantom as described in Ref. [46]; and (iii) the standard Siemens phantom—a cylindrical container filled with nickel sulfate hexahydrate and sodium chloride solution—used on the 3T system.. In the head-shaped phantom setup, the metamaterial was positioned at the back of the "head" to mimic a realistic in-vivo scan. In all other setups, the metamaterial was placed on top of the phantom.

The in-vivo study was approved by the Institutional Review Board of Rabin Medical Center (Petah Tikva, Israel), and all scans were performed after obtaining written informed consent. The in-vivo scans included $B_1^+$ maps acquired with two metamaterial configurations (AD setup 1 and MM2, shown in Fig. 2), as well as a reference $B_1^+$ map without a metamaterial. Additionally, three MRI sequences were acquired using the MM2 configuration to assess clinically relevant applications. The parameters of these three scans are summarized below. (1) a non-contrast-enhanced time-of-flight (TOF) sequence for magnetic resonance angiography (MRA) - FOV 220×176×33 mm³, spatial resolution 0.3×0.3×0.3 mm³, 2 slabs, bandwidth per pixel 120 Hz, TR/TE 27/5.61 ms, flip angle 15°, acceleration factor ×3, scan duration 6:44 minutes.

(2) a high-resolution 3D $T_2$-weighted sequence for detecting structural brain pathology: SPACE sequence with FOV =220 × 150 × 170 mm$^3$, spatial resolution = 0.67 × 0.67 × 0.67 mm$^3$ , TR/TE = 4500/118 ms, acceleration factor ×6, scan duration 7:21 min.

(3) a multi-echo multi-slice gradient-echo acquisition for generating quantitative $T_2^*$ relaxation time maps: FOV =210 × 240 × 12 cm$^3$, spatial resolution = 0.75 × 0.75 × 0.8 mm$^3$, acceleration factor ×2, TR = 2.8 s, five echo times TE=6.44ms, ΔTE=3.3 ms, and flip angle = 55°.

## Acknowledgments

We are grateful to the Weizmann Institute's MRI technician team – Dr. Edna Furman-Haran, E. Tegareh and N. Oshri - for assistance in the human imaging scans. This work was supported by a grant from the Israel Science Foundation (ISF grant No. 1252/23). Dr. R. Schmidt's lab research was generously supported by Mike and Valeria Rosenbloom Center for Research on Positive Neuroscience and Joyce Eisenberg Keefer and Mel Keefer Career Development Chair for New Scientists.

## Competing interests

No conflict of interest.

References:

1. Dumoulin, S. O., Fracasso, A., van der Zwaag, W., Siero, J. C. W. & Petridou, N. Ultra-high field MRI: Advancing systems neuroscience towards mesoscopic human brain function. *NeuroImage* **168**, 345–357 (2018).

2. Webb, A., Shchelokova, A., Slobozhanyuk, A., Zivkovic, I. & Schmidt, R. Novel materials in magnetic resonance imaging: high permittivity ceramics, metamaterials, metasurfaces and artificial dielectrics. *Magn Reson Mater Phy* **35**, 875–894 (2022).

3. Wu, K., Zhu, X., Anderson, S. W. & Zhang, X. Wireless, customizable coaxially shielded coils for magnetic resonance imaging. *Sci. Adv.* **10**, eadn5195 (2024).

4. Schmidt, R., Slobozhanyuk, A., Belov, P. & Webb, A. Flexible and compact hybrid metasurfaces for enhanced ultra high field in vivo magnetic resonance imaging. *Sci Rep* **7**, 1678 (2017).

5. Chi, Z. *et al.* Adaptive Cylindrical Wireless Metasurfaces in Clinical Magnetic Resonance Imaging. *Advanced Materials* **33**, 2102469 (2021).

6. Puchnin, V. *et al.* Metamaterial inspired wireless coil for clinical breast imaging. *Journal of Magnetic Resonance* **322**, 106877 (2021).

7. Shchelokova, A. V. *et al.* Volumetric wireless coil based on periodically coupled split-loop resonators for clinical wrist imaging. *Magnetic Resonance in Med* **80**, 1726–1737 (2018).

8. Motovilova, E., Sandeep, S., Hashimoto, M. & Huang, S. Y. Water-Tunable Highly Sub-Wavelength Spiral Resonator for Magnetic Field Enhancement of MRI Coils at 1.5 T. *IEEE Access* **7**, 90304–90315 (2019).

9. Hurshkainen, A. A. *et al.* EBG metasurfaces for MRI application. in *2016 IEEE Radio and Antenna Days of the Indian Ocean (RADIO)* 1–2 (IEEE, Reunion, France, 2016). doi:10.1109/RADIO.2016.7772010.

10. Nurzed, B., Saha, N., Millward, J. M. & Niendorf, T. 3D Metamaterials Facilitate Human Cardiac MRI at 21.0 Tesla: A Proof-of-Concept Study. *Sensors* **25**, 620 (2025).

11. Van de Moortele, P.-F. *et al.* B1 destructive interferences and spatial phase patterns at 7 T with a head transceiver array coil. *Magn. Reson. Med.* **54**, 1503–1518 (2005).

12. Collins, C. M., Li, S. & Smith, M. B. SAR and B1 field distributions in a heterogeneous human head model within a birdcage coil. *Magn. Reson. Med.* **40**, 847–856 (1998).

13. Yang, Q. X. *et al.* Analysis of wave behavior in lossy dielectric samples at high field. *Magnetic Resonance in Med* **47**, 982–989 (2002).

14. Guérin, B., Gebhardt, M., Cauley, S., Adalsteinsson, E. & Wald, L. L. Local specific absorption rate (SAR), global SAR, transmitter power, and excitation accuracy trade-offs in low flip-angle parallel transmit pulse design. *Magnetic Resonance in Med* **71**, 1446–1457 (2014).

15. Deniz, C. M. Parallel Transmission for Ultrahigh Field MRI. *Topics in Magnetic Resonance Imaging* **28**, 159–171 (2019).

16. Steensma, B. R. *et al.* An 8-channel Tx/Rx dipole array combined with 16 Rx loops for high-resolution functional cardiac imaging at 7 T. *Magn Reson Mater Phy* **31**, 7–18 (2018).

17. Adriany, G. *et al.* Evaluation of a 16-Channel Transmitter for Head Imaging at 10.5T. in *2019 International Conference on Electromagnetics in Advanced Applications (ICEAA)* 1171–1174 (IEEE, Granada, Spain, 2019). doi:10.1109/ICEAA.2019.8879131.

18. Ertürk, M. A., Raaijmakers, A. J. E., Adriany, G., Uğurbil, K. & Metzger, G. J. A 16-channel combined loop-dipole transceiver array for 7 T esla body MRI. *Magnetic Resonance in Med* **77**, 884–894 (2017).

19. Sengupta, S. *et al.* A Specialized Multi-Transmit Head Coil for High Resolution fMRI of the Human Visual Cortex at 7T. *PLoS ONE* **11**, e0165418 (2016).

20. Lagore, R. L. *et al.* An 8-dipole transceive and 24-loop receive array for non-human primate head imaging at 10.5 T. *NMR in Biomedicine* **34**, e4472 (2021).

21. Brink, W. M. & Webb, A. G. High permittivity pads reduce specific absorption rate, improve $B_1$ homogeneity, and increase contrast-to-noise ratio for functional cardiac MRI at 3 T. *Magnetic Resonance in Med* **71**, 1632–1640 (2014).

22. van Gemert, J., Brink, W., Webb, A. & Remis, R. High-permittivity pad design tool for 7T neuroimaging and 3T body imaging. *Magn Reson Med* **81**, 3370–3378 (2019).

23. Yang, Q. X. *et al.* Radiofrequency field enhancement with high dielectric constant (HDC) pads in a receive array coil at 3.0T. *Magnetic Resonance Imaging* **38**, 435–440 (2013).

24. Luo, W. *et al.* Permittivity and performance of dielectric pads with sintered ceramic beads in MRI: early experiments and simulations at 3 T. *Magnetic Resonance in Med* **70**, 269–275 (2013).

25. Yang, Q. X. *et al.* Manipulation of image intensity distribution at 7.0 T: Passive RF shimming and focusing with dielectric materials. *Magnetic Resonance Imaging* **24**, 197–202 (2006).

26. Yang, Q. X. *et al.* Reducing SAR and enhancing cerebral signal-to-noise ratio with high permittivity padding at 3 T: Reducing SAR and Enhancing SNR. *Magn. Reson. Med.* **65**, 358–362 (2011).

27. Webb, A. G. Dielectric materials in magnetic resonance. *Concepts Magn. Reson.* **38A**, 148–184 (2011).

28. Brink, W. M., van der Jagt, A. M. A., Versluis, M. J., Verbist, B. M. & Webb, A. G. High Permittivity Dielectric Pads Improve High Spatial Resolution Magnetic Resonance Imaging of the Inner Ear at 7 T: *Investigative Radiology* **49**, 271–277 (2014).

29. Vaidya, M. V. *et al.* Improved detection of fMRI activation in the cerebellum at 7T with dielectric pads extending the imaging region of a commercial head coil. *Magnetic Resonance Imaging* **48**, 431–440 (2018).

30. Miranda, V., Ruello, G. & Lattanzi, R. A theoretical framework to investigate the effect of high permittivity materials in MRI using anatomy-mimicking cylinders. *Magnetic Resonance in Med* **92**, 416–429 (2024).

31. Carluccio, G. & Collins, C. M. High-permittivity pads to enhance SNR and transmit efficiency in MRI of the heart at 7T: a simulation study. *Magn Reson Mater Phy* **35**, 903–909 (2022).

32. Schmidt, R. & Webb, A. Metamaterial Combining Electric- and Magnetic-Dipole-Based Configurations for Unique Dual-Band Signal Enhancement in Ultrahigh-Field Magnetic Resonance Imaging. *ACS Appl. Mater. Interfaces* **9**, 34618–34624 (2017).

33. Stoja, E. *et al.* Improving magnetic resonance imaging with smart and thin metasurfaces. *Sci Rep* **11**, 16179 (2021).

34. Alipour, A. *et al.* Improvement of magnetic resonance imaging using a wireless radiofrequency resonator array. *Sci Rep* **11**, 23034 (2021).

35. Maurya, S. K. & Schmidt, R. A Metamaterial-like Structure Design Using Non-uniformly Distributed Dielectric and Conducting Strips to Boost the RF Field Distribution in 7 T MRI. *Sensors* **24**, 2250 (2024).

36. Vergara Gomez, T. S. *et al.* Hilbert fractal inspired dipoles for passive RF shimming in ultra-high field MRI. *Photonics and Nanostructures - Fundamentals and Applications* **48**, 100988 (2022).

37. Vorobyev, V. *et al.* Improving homogeneity in abdominal imaging at 3 T with light, flexible, and compact metasurface. *Magnetic Resonance in Med* **87**, 496–508 (2022).

38. Koloskov, V., Brink, W. M., Webb, A. G. & Shchelokova, A. Flexible metasurface for improving brain imaging at 7T. *Magnetic Resonance in Med* **92**, 869–880 (2024).

39. Jacobs, P. S. *et al.* In vivo B1+ enhancement of calf MRI at 7 T via optimized flexible metasurfaces. *Magnetic Resonance in Med* **92**, 1277–1289 (2024).

40. Maurya, S. K. & Schmidt, R. Shaping the RF Transmit Field in 7T MRI Using a Nonuniform Metasurface Constructed of Short Conducting Strips. *ACS Appl. Mater. Interfaces* **16**, 47284–47293 (2024).

41. Haines, K., Smith, N. B. & Webb, A. G. New high dielectric constant materials for tailoring the B 1 + distribution at high magnetic fields. *Journal of Magnetic Resonance* **203**, 323–327 (2010).

42. Wu, X. *et al.* Human Connectome Project-style resting-state functional MRI at 7 Tesla using radiofrequency parallel transmission. *NeuroImage* **184**, 396–408 (2019).

43. Brink, W. M., Versluis, M. J., Peeters, J. M., Börnert, P. & Webb, A. G. Passive radiofrequency shimming in the thighs at 3 Tesla using high permittivity materials and body coil receive uniformity correction. *Magnetic Resonance in Med* **76**, 1951–1956 (2016).

44. Koolstra, K., Börnert, P., Brink, W. & Webb, A. Improved image quality and reduced power deposition in the spine at 3 T using extremely high permittivity materials. *Magnetic Resonance in Med* **79**, 1192–1199 (2018).

45. Zhou, J., Zhang, L., Tuttle, G., Koschny, T. & Soukoulis, C. M. Negative index materials using simple short wire pairs. *Phys. Rev. B* **73**, 041101 (2006).

46. Jona, G., Furman-Haran, E. & Schmidt, R. Realistic head-shaped phantom with brain-mimicking metabolites for 7 T spectroscopy and spectroscopic imaging. *NMR in Biomedicine* **34**, (2021).

47. Gosselin, M.-C. *et al.* Development of a new generation of high-resolution anatomical models for medical device evaluation: the Virtual Population 3.0. *Phys. Med. Biol.* **59**, 5287–5303 (2014).

48. Steensma, B. *et al.* Introduction of the snake antenna array: Geometry optimization of a sinusoidal dipole antenna for 10.5T body imaging with lower peak SAR. *Magnetic Resonance in Med* **84**, 2885–2896 (2020).

Supplementary Materials for

# Modular Metamaterials for Adaptive MRI Signal Control: Combining Dipole Arrays with Hexagon-based Artificial Dielectrics

Santosh Kumar Maurya [1,2], Ilan Goldberg [3], Rita Schmidt [1,2]

[1]Department of Brain Sciences, Weizmann Institute of Science, Israel

[2]The Azrieli National Institute for Human Brain Imaging and Research, Weizmann Institute of Science

[3]Department of Neurology, Rabin Medical Center, Israel

**This PDF file includes:**

Figures S1 to S5

## S1. Phantom imaging

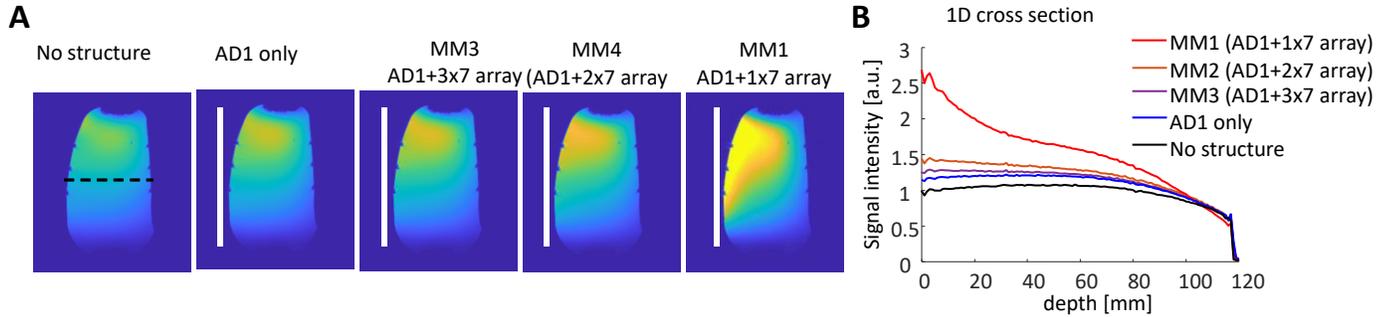

**Figure S1.** Phantom imaging at 7T using modular metamaterial configurations, including AD1 alone, MM3, MM2, and MM1. A) Magnitude images for each setup, B) 1D signal profile along the line indicated in the first image. The white box overlay shows the location of the metamaterial. The maximal intensity increase in the central line was 1.2, 1.3, 1.5, 2.8 for AD1, MM3, MM2, and MM1, respectively. Assuming GRE scan intensity increases approximately with $B_1^2$, the maximal $B_1^+$ enhancement observed was 10%, 14%, 22% and 70%, respectively.

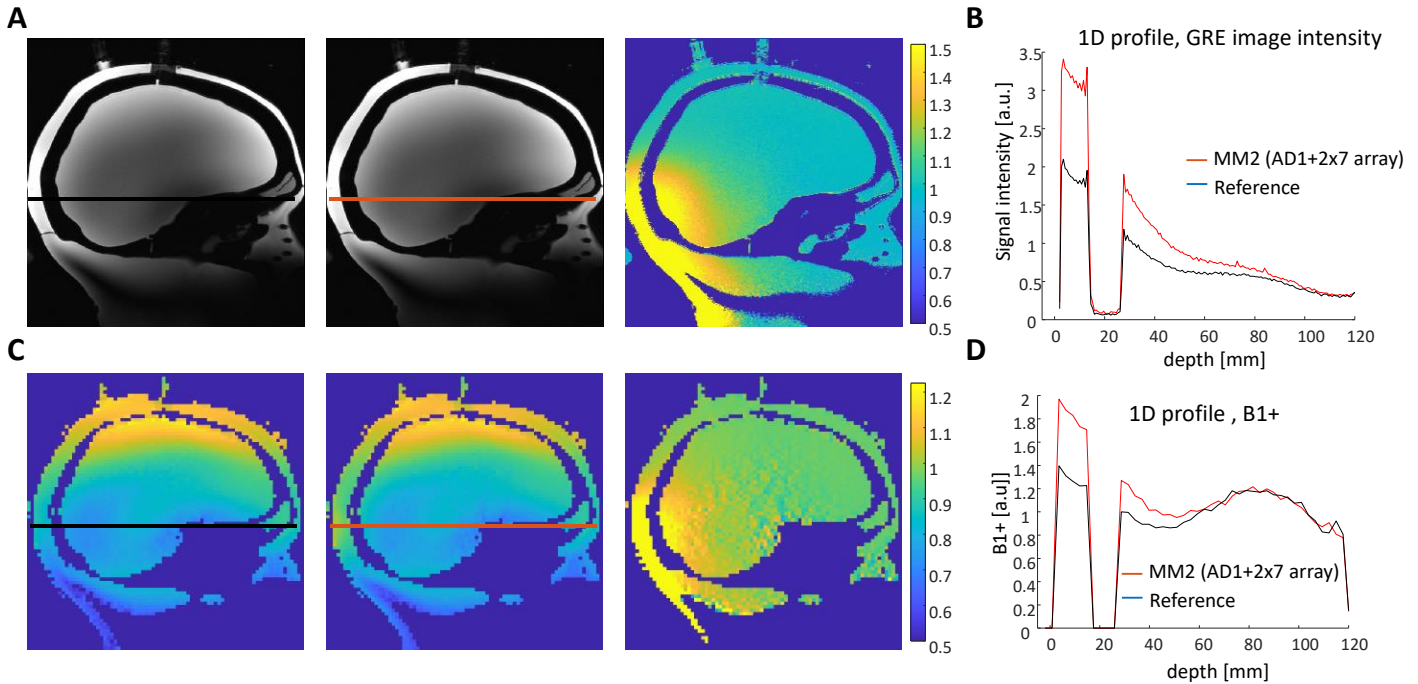

**Figure S2.** 3D head-shaped phantom imaging at 7T with the MM2 metamaterial configuration. A, B) MRI magnitude images acquired without and with the metamaterial, respectively. C, D) Corresponding $B_1^+$ maps without and with the metamaterial. The maximal intensity increase in the illustrated line (B) was 1.65 and maximal $B_1^+$ increase was 1.3. Ratio maps of the magnitude and $B_1^+$ images are shown in panels A and B, respectively.

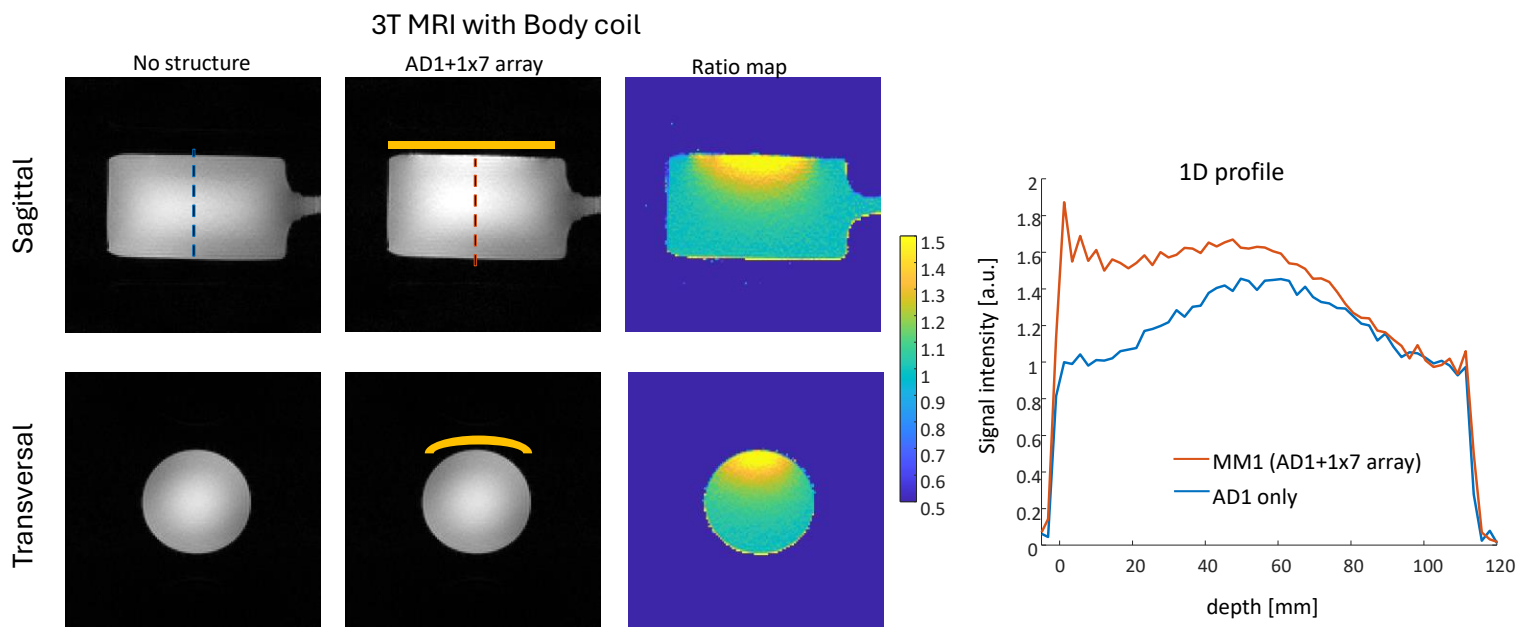

**Figure S3.** Phantom imaging at 3T with the MM1 configuration (AD1 + 1×7 dipole array). Sagittal and transverse images acquired without and with the metamaterial are shown, along with corresponding ratio maps. A 1D signal profile along the line indicated in the images is presented on the right.

## S2. EM simulations with virtual human model

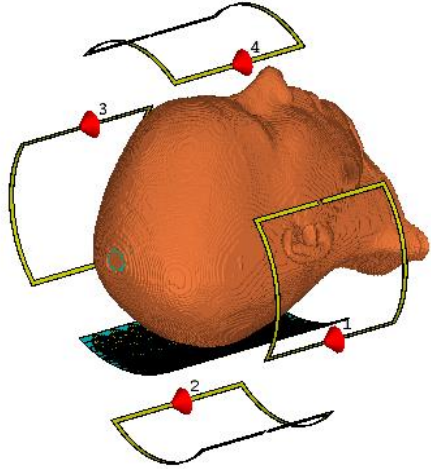

**Figure S4.** Schematic view of the setup used in EM simulations, including the 4-port RF coil, the metamaterial positioned at the back of the head, and the human head model.

## S3. Metamaterial configurations

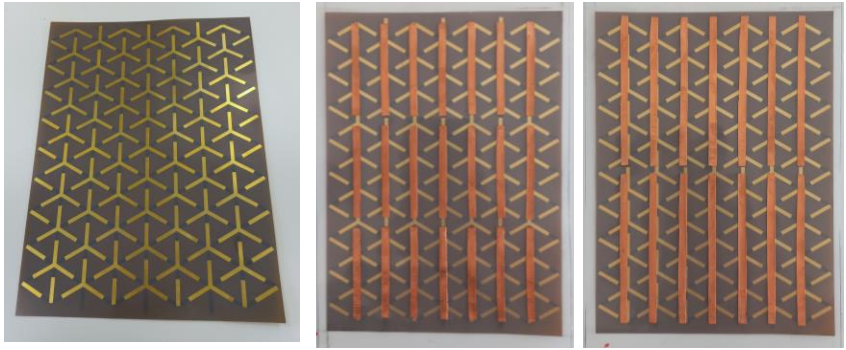

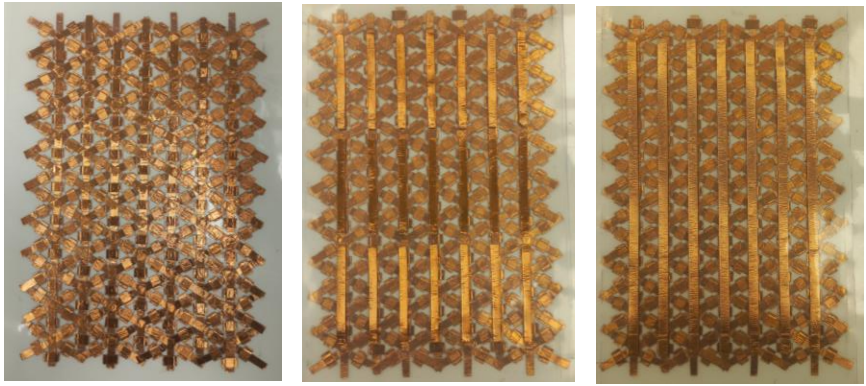

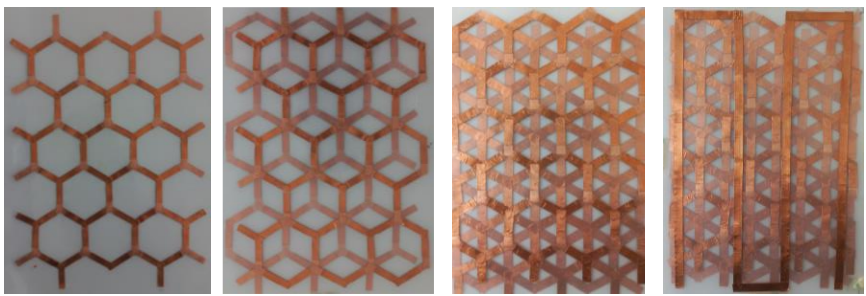

**Figure S5.** Photographs of the fabricated metamaterial configurations. A) Fig. 3 Setup II implemented using flexible PCB. From left to right: artificial dielectric only, with 3×7 dipole array, and with 2×7 dipole array. B) Fig. 3 Setup I low-cost implementation. From left to right: artificial dielectric only, with 3×7 dipole array, and with 1×7 dipole array. C) Fig. 1C multilayer configuration. From left to right: single layer, dual-layer, triple-layer, and triple-layer with snake antenna.